\documentclass[twocolumn,showpacs,preprintnumbers,amsmath,amssymb]{revtex4}

\usepackage{graphicx}
\usepackage{dcolumn}
\usepackage{bm}

\begin{document}

\title{Unified Parametrization of Quark and Lepton Mixing Matrices}

\author{Nan Li}
\author{Bo-Qiang Ma}\altaffiliation{Corresponding author}\email{mabq@phy.pku.edu.cn}
\affiliation{School of Physics, Peking University, Beijing 100871,
China}

\begin{abstract}
We present a unified parametrization of quark and lepton mixing
matrices. By using some simple relations between the mixing angles
of quarks and leptons, i.e., the quark-lepton complementarity, we
parametrize the lepton mixing matrix with the Wolfenstein
parameters $\lambda$ and $A$ of the quark mixing matrix.
It is shown that the Wolfenstein parameter $\lambda$ can measure
both the deviation of the quark mixing matrix from the unit
matrix, and the deviation of the lepton mixing matrix from the
exactly bimaximal mixing pattern.
\end{abstract}


\maketitle

{\it Introduction}---To describe the behaviors of quarks and
leptons in a grand unified theory (GUT) is one of main goals of
particle physics. Among all the characters of quarks and leptons,
the mixing between different generations is one of the fundamental
problems. Before more underlying theory of the origin of the
mixing is found, to parametrize the quark mixing (CKM)
matrix~\cite{ckm} and the lepton mixing (PMNS) matrix~\cite{pmns}
phenomenologically is the first step to understand this problem.
However, these two mixing matrices were parametrized in isolated
ways, with the parameters in these two mixing matrices being
uncorrelated with each other. The purpose of this work is to show
that one can parametrize the quark and lepton mixing matrices in a
unified way by adopting some simple relations between the mixing
angles of quarks and leptons
\begin{eqnarray}
&&\theta_{23}+\theta_{23}^{\prime}(\theta_{\mathrm{atm}})=\frac{\pi}{4},\nonumber \\
&&\theta_{31} \sim \theta_{31}^{\prime}(\theta_{\mathrm{chz}}),\nonumber\\
&&\theta_{12}(\theta_{\mathrm{C}})+\theta_{12}^{\prime}
(\theta_{\mathrm{sol}})=\frac{\pi}{4}, \label{eq.correlation}
\end{eqnarray}
where $\theta_{ij}$ and $\theta^{\prime}_{ij}$ (for $i,j=1, 2, 3$)
are the mixing angles of the $i$ and $j$ generations of the CKM
matrix and the PMNS matrix ($\theta_{12}$ is the Cabibbo mixing
angle $\theta_\mathrm{C}$). These relations, which have been
suggested by Raidal~\cite{raidal} as a support of the grand
quark-lepton unification or certain quark-lepton symmetry, are in
perfect agreement with experimental data (for example,
$\theta_{\mathrm{C}}=12.9^{\circ}$ and
$\theta_{\mathrm{sol}}=32.6^{\circ}$ at the best fit points, and
$\theta_{\mathrm{C}}+\theta_{\mathrm{sol}}=45.5^{\circ}$). The
third numerical correlation has been pointed out by
Smirnov~\cite{Smirnov}, and is called the quark-lepton
complementarity (QLC)~\cite{Complementarity}.

From these relations, we can find that the mixing angles of quarks
and leptons are not independent of each other. So we can get the
trigonometric functions of the mixing angles of leptons in terms
of these of quarks, and link the parameters of the PMNS matrix
with these of the CKM matrix. Therefore, we can parametrize the
PMNS matrix with the parameters of the CKM matrix, and express the
CKM and the PMNS matrices in a same framework.

{\it The quark and lepton mixing matrices}---Both quark and lepton
mixing matrices can be written as
\begin{equation}
    \left(
        \begin{array}{ccc}
            c_{31}c_{12} & c_{31}s_{12} & s_{31}e^{-i\delta}\\
           -c_{23}s_{12}-s_{23}s_{31}c_{12}e^{i\delta} & c_{23}c_{12}-s_{23}s_{31}s_{12}e^{i\delta} & s_{23}c_{31}\\
            s_{23}s_{12}-c_{23}s_{31}c_{12}e^{i\delta} & -s_{23}c_{12}-c_{23}s_{31}s_{12}e^{i\delta} & c_{23}c_{31}\\
        \end{array}
        \right),\label{eq.mixing matrix}
\end{equation}
where $s_{ij}=\sin\theta_{ij}$, $c_{ij}=\cos\theta_{ij}$ (for
$i,j=1, 2, 3$), 
and $\delta$ is the $CP$-violating phase. Altogether there are
four parameters in the mixing matrix, describing both the real and
the imaginary parts of the mixing matrix.

For the CKM matrix $V$, the best fit values of the three mixing
angles are $\theta_{12} (\theta_{\mathrm{C}}) =12.9^{\circ}$,
$\theta_{23}=2.4^{\circ}$, and
$\theta_{31}=0.2^{\circ}$~\cite{pdg}, and we can find that all the
three mixing angles are not large. So the CKM matrix is a small
deviation from the unit matrix, and it can be parametrized
as~\cite{wol}
\begin{equation}
    V=\left(
        \begin{array}{ccc}
             1-\frac{1}{2}\lambda^2 & \lambda & A\lambda^3(\rho-i\eta)\\
             -\lambda & 1-\frac{1}{2}\lambda^2 & A\lambda^2\\
             A\lambda^3(1-\rho-i\eta) & -A\lambda^2 & 1\\
        \end{array}
        \right),\label{eq.wol}
\end{equation}
where $\lambda$ measures the strength of the deviation of $V$ from
the unit matrix
($\lambda=\sin\theta_{\mathrm{C}}=0.2243\pm0.0016$), and $A$,
$\rho$ and $\eta$ are the other three parameters, with the best
fit values $A=0.82$, $\rho=0.20$ and $\eta=0.33$~\cite{pdg}.

However, for the PMNS matrix $U$, the situation is quite different
from the CKM matrix. With the help of the experimental data from
KamLAND~\cite{Kam}, SNO~\cite{sno}, K2K~\cite{K2K},
Super-Kamiokande~\cite{SUPER} and CHOOZ~\cite{Chz} experiments, we
know that the mixing angles of leptons are not as small as those
of quarks~\cite{raidal},
\begin{eqnarray}
&&\sin^{2}2\theta_{\mathrm{atm}}=1.00\pm0.05, \nonumber\\
&&\sin^{2}2\theta_{\mathrm{chz}}=0\pm0.065, \nonumber\\
&&\tan^2\theta_{\mathrm{sol}}=0.41\pm0.05,\label{eq.mixing angles}
\end{eqnarray}
where $\theta_{\mathrm{atm}}$, $\theta_{\mathrm{chz}}$, and
$\theta_{\mathrm{sol}}$ are the mixing angles of atmospheric,
CHOOZ and solar neutrino oscillations, and we have
$\theta_{\mathrm{atm}}=45.0^{\circ}\pm6.5^{\circ}$,
$\theta_{\mathrm{chz}}=0^{\circ}\pm7.4^{\circ}$ and
$\theta_{\mathrm{sol}}=32.6^{\circ}\pm1.6^{\circ}$. So the
numerical relations in Eq.~(\ref{eq.correlation}) are satisfied to
a good degree of accuracy.

Therefore, we can get the PMNS matrix and find that almost all the
non-diagonal elements of the PMNS matrix are large. According to
the results of the global analysis of the neutrino oscillation
experimental data, the elements of the modulus of the PMNS matrix
are summarized as~\cite{garcia}
\begin{equation}
    |U|=\left(
        \begin{array}{ccc}
             0.77-0.88 & 0.47-0.61 & <0.20\\
             0.19-0.52 & 0.42-0.73 & 0.58-0.82\\
             0.20-0.53 & 0.44-0.74 & 0.56-0.81\\
        \end{array}
        \right).\label{eq.modulus}
\end{equation}
We can see from Eq.~(\ref{eq.modulus}) that the PMNS matrix
deviates from the unit matrix significantly, but it is quite near
the bimaximal mixing pattern, which reads
\begin{equation}
\left(
        \begin{array}{ccc}
            {\sqrt{2}}/{2} & {\sqrt{2}}/{2} & 0 \\
            -1/2 & 1/2 & {\sqrt{2}}/{2} \\
            1/2 & -1/2 & {\sqrt{2}}/{2}
        \end{array} \right).\label{eq.bimaximal}
\end{equation}

So, for the parametrization of the PMNS matrix, it is unpractical
to imitate the Wolfenstein parametrization of the CKM matrix
indiscriminately. The parametrizations of the PMNS matrix on the
basis of the bimaximal mixing pattern have been discussed by
Rodejohann~\cite{Rodejohann} and us~\cite{li}. However, these
parametrizations only concern about the experimental data of
leptons, without taking into account their relations with quarks.
Thus, the CKM and the PMNS matrices are parametrized irrelevantly,
and the parameters in them are not correlated with each other.
However, with the relations in Eq.~(\ref{eq.correlation}), we can
parametrize the quark and lepton mixing matrices with correlated
parameters.

{\it Parametrization of the PMNS matrix}---In Wolfenstein
parametrization of the CKM matrix, we have (to the order of
$\lambda^3$)
\begin{eqnarray}\nonumber
&&\sin\theta_{12}=\lambda,\quad
\cos\theta_{12}=1-\frac{1}{2}\lambda^2,\nonumber\\
&&\sin\theta_{23}=A\lambda^2, \quad \cos\theta_{23}=1,\nonumber\\
&&\sin\theta_{31}e^{-i\delta}=A\lambda^3(\rho-i\eta), \quad
\cos\theta_{31}=1.\label{eq.wolfenstein}
\end{eqnarray}

For the case of leptons, using Eq.~(\ref{eq.correlation}), we have
(to the order of $\lambda^3$)
\begin{eqnarray}
&&\sin\theta_{12}^{\prime}=\sin(\frac{\pi}{4}-\theta_{12})=\frac{\sqrt{2}}{2}(1-\lambda-\frac{1}{2}\lambda^2), \nonumber\\
&&\cos\theta_{12}^{\prime}=\frac{\sqrt{2}}{2}(1+\lambda-\frac{1}{2}\lambda^2), \nonumber \\
&&\sin\theta_{23}^{\prime}=\frac{\sqrt{2}}{2}(1-A\lambda^2),
\nonumber \\&&
\cos\theta_{23}^{\prime}=\frac{\sqrt{2}}{2}(1+A\lambda^2), \nonumber \\
&&\sin\theta_{31}^{\prime}e^{-i\delta^{\prime}}=A\lambda^3(\zeta-i\xi),
\nonumber \\&& \cos\theta_{31}^{\prime}=1,\label{eq.sincos}
\end{eqnarray}
where $A$ and $\lambda$ are just the Wolfenstein parameters of the
CKM matrix. So the CKM and the PMNS matrices have only one set of
parameters in this unified parametrization. Because there are
totally four angles in the mixing matrix (three mixing angles and
one $CP$-violating phase angle), and only two precise numerical
relations are known (Eq.~(\ref{eq.correlation})), we have to
introduce another two new parameters $\zeta$ and $\xi$ to describe
the PMNS matrix fully.

In Eq.~(\ref{eq.sincos}), we set
$\sin\theta_{31}^{\prime}e^{-i\delta^{\prime}}=A\lambda^3(\zeta-i\xi)$.
Because of the inaccurate experimental data of neutrino
oscillations, we have not fixed the value of $|U_{e3}|$, and only
known its upper bound~\cite{garcia}. Therefore, we may also set
$\sin\theta_{31}^{\prime}e^{-i\delta^{\prime}}=A\lambda^2(\zeta^{\prime}-i\xi^{\prime})$.
Choosing which of them is to be determined by the future
experimental data, and we discuss these two cases here,
respectively.

{\it Case 1}:
$\sin\theta_{31}^{\prime}e^{-i\delta^{\prime}}=A\lambda^3(\zeta-i\xi)$.

Substituting Eq.~(\ref{eq.sincos}) into Eq.~(\ref{eq.mixing
matrix}), we can get the PMNS matrix as

\begin{widetext}
\begin{eqnarray}\nonumber
   U&=&\left(
        \begin{array}{ccc}
            \frac{\sqrt{2}}{2}(1+\lambda-\frac{1}{2}\lambda^2) & \frac{\sqrt{2}}{2}(1-\lambda-\frac{1}{2}\lambda^2) & A\lambda^3(\zeta-i\xi) \\
            -\frac{1}{2}[1-\lambda+(A-\frac{1}{2})\lambda^2-A\lambda^3(1-\zeta-i\xi)] & \frac{1}{2}[1+\lambda+(A-\frac{1}{2})\lambda^2+A\lambda^3(1-\zeta-i\xi)] & \frac{\sqrt{2}}{2}(1-A\lambda^2) \\
            \frac{1}{2}[1-\lambda-(A+\frac{1}{2})\lambda^2+A\lambda^3(1-\zeta-i\xi)] & -\frac{1}{2}[1+\lambda-(A+\frac{1}{2})\lambda^2-A\lambda^3(1-\zeta-i\xi)] & \frac{\sqrt{2}}{2}(1+A\lambda^2)
        \end{array} \right)\\&=&
\left(
        \begin{array}{ccc}
            \frac{\sqrt{2}}{2} & \frac{\sqrt{2}}{2} & 0 \\
            -\frac{1}{2} & \frac{1}{2} & \frac{\sqrt{2}}{2} \\
            \frac{1}{2} & -\frac{1}{2} & \frac{\sqrt{2}}{2}
        \end{array} \right)+\lambda
\left(
        \begin{array}{ccc}
             \frac{\sqrt{2}}{2} & -\frac{\sqrt{2}}{2} & 0 \\
            \frac{1}{2} & \frac{1}{2} & 0\\
           - \frac{1}{2} & -\frac{1}{2} & 0
        \end{array} \right)+\lambda^{2}
\left(
        \begin{array}{ccc}
            -\frac{\sqrt{2}}{4} & -\frac{\sqrt{2}}{4} & 0 \\
            -\frac{1}{2}(A-\frac{1}{2}) & \frac{1}{2}(A-\frac{1}{2}) & -\frac{\sqrt{2}}{2}A \\
            -\frac{1}{2}(A+\frac{1}{2}) & \frac{1}{2}(A+\frac{1}{2}) & \frac{\sqrt{2}}{2}A
        \end{array} \right)\nonumber\\&&+\lambda^{3}
\left(
        \begin{array}{ccc}
            0 & 0 & A(\zeta-i\xi) \\
            \frac{1}{2}A(1-\zeta-i\xi) & \frac{1}{2}A(1-\zeta-i\xi) & 0\\
            \frac{1}{2}A(1-\zeta-i\xi) & \frac{1}{2}A(1-\zeta-i\xi) & 0
        \end{array} \right)+\mathcal{O}(\lambda^4).\label{eq.3ci}
\end{eqnarray}
\end{widetext}

Now we give some discussion about Eq.~(\ref{eq.3ci}):

(1). The bimaximal mixing pattern is derived naturally as the
leading-order approximation. However, it is chosen as the basis
for the expansion of the PMNS matrix by hand
before~\cite{Rodejohann,li}. So we can even freely choose other
bases for the parametrization of the PMNS matrix (for example, to
parametrize the PMNS matrix in the tri-bimaximal mixing
pattern~\cite{linan}). Now we find that the leading-order term of
the PMNS matrix must be the bimaximal mixing pattern as long as we
accept the numerical relations in Eq.~(\ref{eq.correlation}).

(2). The Wolfenstein parameter $\lambda$ can characterize both the
deviation of the CKM matrix from the unit matrix (see
Eq.~(\ref{eq.wol})), and the deviation of the PMNS matrix from the
exactly bimaximal mixing pattern (see the next-to-leading-order
term in Eq.~(\ref{eq.3ci})). However, in the previous
work~\cite{Rodejohann,li}, $\lambda$ in the PMNS matrix is
introduced independently, without considering its relation with
the Wolfenstein parameter $\lambda$ in the CKM matrix. Now, we can
see that in this unified parametrization these two different
deviations of quarks and leptons are correlated essentially, and
can be measured by only one single parameter $\lambda$, as Raidal
pointed out~\cite{raidal}.

(3). Comparing with the parametrizations in bimaxiaml mixing
pattern~\cite{Rodejohann,li}, we can see that this unified
parametrization is equivalent to them to the leading and
next-to-leading orders. In~\cite{Rodejohann}, the elements of the
PMNS matrix are set to be $U_{e2}=\frac{\sqrt{2}}{2}(1-\lambda)$,
$U_{e3}=A\lambda^ne^{-i\delta}$, and
$U_{\mu3}=\frac{\sqrt{2}}{2}(1-B\lambda^m)$. If we let
$B\rightarrow A$ and fix $m$ to be 2, and $n$ to be 3, we can find
that the parametrization in~\cite{Rodejohann} is just the unified
parametrization here. Similarly, in~\cite{li},
$U_{e1}=\frac{\sqrt{2}}{2}+\lambda$, $U_{e3}=b\lambda^2$, and
$U_{\mu3}=\frac{\sqrt{2}}{2}+a\lambda^2$. If we rescale
$\lambda\rightarrow\frac{\sqrt{2}}{2}\lambda$ and
$a\rightarrow-\sqrt{2}A$, we can find that the first two terms of
the expansion in~\cite{li} are just the same as Eq.~(\ref{eq.3ci})
(in~\cite{li}, $U_{e3}$ is set to be $b\lambda^2$, not
$b\lambda^3$, but this only affects the terms of higher orders).
So the parametrizations in~\cite{Rodejohann,li} have been
rederived as the natural results in this unified parametrization.

(4). The range of $\lambda$ in~\cite{li} is calculated in detail,
$0.08<\lambda<0.17$. Now, in this unified parametrization,
$\lambda$ here is just the Wolfenstein parameter of the CKM
matrix, $\lambda=\sin\theta_{\mathrm{C}}=0.2243$. As discussed in
(3), if we rescale $\lambda$, and divide it by ${\sqrt{2}}$, we
get $\lambda=0.1586$. We can see that the value of the rescaled
$\lambda$ is just in the range calculated in~\cite{li}. So this
unified parametrization is reasonable compared with the
experimental data.

(5). The values of $\rho$ and $\eta$ in the CKM matrix have been
measured by many experiments~\cite{pdg}, and the typical values
are $\rho=0.20$ and $\eta=0.33$. On the contrary, the inaccuracy
of the current experimental data of neutrinos makes it difficult
to fix the values of the elements of the PMNS matrix to a very
good degree of accuracy. So the values of $\zeta$ and $\xi$ have
not been determined by now. At present, the best fit point of
$\sin^2\theta^{\prime}_{31}$ is 0.006~\cite{al}, so we have
$A\lambda^3\sqrt{\zeta^2+\xi^2} \sim 0.077$, and
$\sqrt{\zeta^2+\xi^2} \sim 8.2$. Therefore, both $\zeta$ and $\xi$
are of $\mathcal{O}(1)$.

Furthermore, $\zeta$ and $\xi$ are related with the $CP$-violating
process~\cite{zetaxi}, and the rephasing-invariant measurement of
the lepton $CP$-violation is described by the Jarlskog parameter
$J$~\cite{Ja},
$J=\mbox{Im}(U_{e2}U_{\mu3}U_{e3}^{\ast}U_{\mu2}^{\ast})$. In this
unified parametrization, from Eq.~(\ref{eq.3ci}), $J$ can be
expressed in a simple form (to the order of $\lambda^5$),
\begin{equation}
J=\frac{1}{4}A\lambda^3\xi(1-2\lambda^2)=0.0022\xi.\label{eq.j1}
\end{equation}
We can see from Eq.~(\ref{eq.j1}) that $J$ is only related with
the parameter $\xi$. So if we can observe the lepton
$CP$-violating process in the future neutrinoless $\beta\beta$
decay reaction~\cite{beta}, and can determine the value of $J$,
then the value of $\xi$ can be fixed. And with the more precise
experimental data of $|U_{e3}|$, we can determine the value of
$\zeta$ ultimately. Thus we can get a full understanding of the
structure of the PMNS matrix.

{\it Case 2}:
$\sin\theta_{31}^{\prime}e^{-i\delta^{\prime}}=A\lambda^2(\zeta^{\prime}-i\xi^{\prime})$.

Repeating the former process, we get

\begin{widetext}
\begin{eqnarray}\nonumber
   U&=&\left(
        \begin{array}{ccc}
            \frac{\sqrt{2}}{2}(1+\lambda-\frac{1}{2}\lambda^2) & \frac{\sqrt{2}}{2}(1-\lambda-\frac{1}{2}\lambda^2) & A\lambda^2(\zeta^{\prime}-i\xi^{\prime}) \\
            -\frac{1}{2}\{1-\lambda-[\frac{1}{2}-A(1+\zeta^{\prime}+i\xi^{\prime})]\lambda^2\} & \frac{1}{2}\{1+\lambda-[\frac{1}{2}-A(1-\zeta^{\prime}-i\xi^{\prime})]\lambda^2\} & \frac{\sqrt{2}}{2}(1-A\lambda^2) \\
            \frac{1}{2}\{1-\lambda-[\frac{1}{2}+A(1+\zeta^{\prime}+i\xi^{\prime})]\lambda^2\} & -\frac{1}{2}\{1+\lambda-[\frac{1}{2}+A(1-\zeta^{\prime}-i\xi^{\prime})]\lambda^2\} & \frac{\sqrt{2}}{2}(1+A\lambda^2)
        \end{array} \right)\\&=&
\left(
        \begin{array}{ccc}
            \frac{\sqrt{2}}{2} & \frac{\sqrt{2}}{2} & 0 \\
            -\frac{1}{2} & \frac{1}{2} & \frac{\sqrt{2}}{2} \\
            \frac{1}{2} & -\frac{1}{2} & \frac{\sqrt{2}}{2}
        \end{array} \right)+\lambda
 \left(
        \begin{array}{ccc}
             \frac{\sqrt{2}}{2} & -\frac{\sqrt{2}}{2} & 0 \\
            \frac{1}{2} & \frac{1}{2} & 0\\
           - \frac{1}{2} & -\frac{1}{2} & 0
        \end{array} \right)\nonumber \\&&
        +\lambda^{2}
 \left(
        \begin{array}{ccc}
            -\frac{\sqrt{2}}{4} & -\frac{\sqrt{2}}{4} & A(\zeta^{\prime}-i\xi^{\prime}) \\
            \frac{1}{2}[\frac{1}{2}-A(1+\zeta^{\prime}+i\xi^{\prime})] & -\frac{1}{2}[\frac{1}{2}-A(1-\zeta^{\prime}-i\xi^{\prime})] & -\frac{\sqrt{2}}{2}A\\
            -\frac{1}{2}[\frac{1}{2}+A(1+\zeta^{\prime}+i\xi^{\prime})] & \frac{1}{2}[\frac{1}{2}+A(1-\zeta^{\prime}-i\xi^{\prime})] & \frac{\sqrt{2}}{2}A
        \end{array} \right)+\mathcal{O}(\lambda^3).\label{eq.2ci}
\end{eqnarray}
\end{widetext}

Similar to {\it Case 1}, we can see that:

(1). The bimaximal mixing pattern is derived as the leading-order
term naturally.

(2). The deviation of the CKM matrix from the unit matrix, and the
deviation of the PMNS matrix from the exactly bimaximal mixing
pattern can be characterized by only one parameter $\lambda$.

(3). Parametrizations in~\cite{Rodejohann,li} can be transformed
into this unified parametrization, if we let $B\rightarrow A$ and
fix $m$ and $n$ to be 2 in~\cite{Rodejohann}, and rescale
$\lambda\rightarrow\frac{\sqrt{2}}{2}\lambda$ and
$a\rightarrow-\sqrt{2}A$ in~\cite{li}.

(4). The expressions of the leading-order and
next-to-leading-order terms in {\it Case 2} are the same as those
in {\it Case 1}, because the difference between them is caused by
the introductions of $U_{e3}$ at the second and the third orders.
So the expressions of the first two orders must be the same in
these two cases. Also, $\lambda$ in {\it Case 2} is still
consistent with the range $0.08<\lambda<0.17$ after rescaling.

(5). The Jarlskog parameter $J$ can be expressed now in the form
\begin{equation}
J=\frac{1}{4}A\lambda^2\xi^{\prime}(1-2\lambda^2)=0.0099\xi^{\prime}.\label{eq.j2}
\end{equation}
Similarly, we can fix the value of $\xi^{\prime}$ by observing the
lepton $CP$-violating process, and then can determine the value of
$\zeta^{\prime}$. Now $\sqrt{\zeta^{\prime2}+\xi^{\prime2}} \sim
1.8$, and $\zeta^{\prime}$ and $\xi^{\prime}$ are still of
$\mathcal{O}(1)$. Of course, $\zeta^{\prime}$ and $\xi^{\prime}$
in {\it Case 2} are not the $\zeta$ and $\xi$ in {\it Case 1}, and
they are equivalent to the $\zeta$ and $\xi$ in {\it Case 1} by
rescalings $\zeta^{\prime}\rightarrow\lambda\zeta$ and
$\xi^{\prime}\rightarrow\lambda\xi$.

Then we can see the merits of these two cases. If
$\sin^2\theta_{31}^{\prime} \sim 0.006$ as a preliminary estimate
shows, then {\it Case 2} is preferable, because $\zeta'$ and
$\xi'$ in {\it Case 2} are more close to 1 in magnitude. However,
if $\sin^2 \theta_{31}^{\prime} \sim 0.0001$ or less, then {\it
Case 1} is to be preferred.

{\it Conclusions}---We present a unified parametrization of the
quark and lepton mixing matrices, which is based on the simple
relations between the mixing angles of quarks and leptons.
Although the physical explanation of these relations remains to be
explored, we believe that there must be some deeper principle
behind these elegant correlations, which are in perfect agreement
with the current experimental data.

If the numerical relations in Eq.~(\ref{eq.correlation}) violate a
little, we can maintain the expressions in Eq.~(\ref{eq.sincos}),
and only need to redefine the parameters $\lambda$ and $A$. For
example, we can still set
$\sin\theta_{23}^{\prime}=\frac{\sqrt{2}}{2}(1-A^{\prime}\lambda^{\prime2})$.
Thus, the parameter $\lambda^{\prime}$ and $A^{\prime}$ are not
the same as the Wolfenstein parameters $\lambda$ and $A$, and the
symmetry between the quark and lepton mixing matrices will break
slightly. This is a more general parametrization, and can work
whether Raidal's numerical relations keep or not. However,
$A^{\prime}$ and $\lambda^{\prime}$ in this more general
parametrization are still the Wolfenstein-like parameters. The
forms of $s^{\prime}_{ij}$, $c^{\prime}_{ij}$ and expansion of the
PMNS matrix will still keep invariant, and the leading-order term
is still the bimaximal mixing pattern, only with the transition
$\lambda\rightarrow\lambda^{\prime}$, and $A\rightarrow
A^{\prime}$.

In conclusion, although all sorts of parametrization of the quark
and lepton mixing matrices are not based on deep theoretical
foundation, and applying any of them may not have specific
physical significance, however, it is quite likely that this
unified parametrization does have its advantages. For instance,
the number of the free parameters in this unified parametrization
is fewer than the parametrizations in~\cite{Rodejohann,li}, the
bimaximal mixing pattern as the leading-order term is derived
naturally, and the next-to-leading-order is the same
as~\cite{Rodejohann,li} after rescalings. Also, the Wolfenstein
parameter $\lambda$ can measure both the deviation of the CKM
matrix from the unit matrix, and the deviation of the PMNS matrix
from the bimaximal mixing pattern. So if this unified
parametrization is tested to be consistent with more precise
experimental data in the future, we can get a comprehensive
understanding of the mixings of quarks and leptons, and push
forward the exploration of grand unification.

We are very grateful to Prof.~Xiao-Gang He for
suggestions. This work is partially supported by
National Natural
Science Foundation of China 
and by the Key Grant Project of Chinese Ministry of Education
(No.~305001).


\end{document}